\documentclass[8.5pt,twoside,twocolumn]{article}
\usepackage[super,sort&compress,comma]{natbib} 
\usepackage{mhchem,chemfig}
\usepackage{times,mathptmx}
\usepackage{amssymb}
\usepackage{sectsty}
\usepackage{balance} 

\usepackage{graphicx} 
\usepackage{lastpage}
\usepackage[format=plain,justification=raggedright,singlelinecheck=false,font=small,labelfont=bf,labelsep=space]{caption} 
\usepackage{fancyhdr}
\begin{document}

\thispagestyle{plain}
\fancypagestyle{plain}{
\fancyhead[L]{\includegraphics[height=8pt]{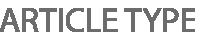}}
\fancyhead[C]{\hspace{-1cm}\includegraphics[height=20pt]{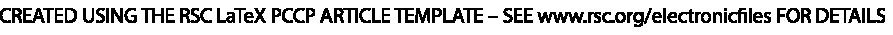}}
\fancyhead[R]{\includegraphics[height=10pt]{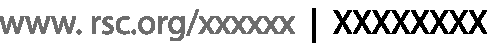}\vspace{-0.2cm}}
\renewcommand{\headrulewidth}{1pt}}
\renewcommand{\thefootnote}{\fnsymbol{footnote}}
\renewcommand\footnoterule{\vspace*{1pt}%
\hrule width 3.4in height 0.4pt \vspace*{5pt}} 
\setcounter{secnumdepth}{5}

\makeatletter 
\def\subsubsection{\@startsection{subsubsection}{3}{10pt}{-1.25ex plus -1ex minus -.1ex}{0ex plus 0ex}{\normalsize\bf}} 
\def\paragraph{\@startsection{paragraph}{4}{10pt}{-1.25ex plus -1ex minus -.1ex}{0ex plus 0ex}{\normalsize\textit}} 
\renewcommand\@biblabel[1]{#1}            
\renewcommand\@makefntext[1]%
{\noindent\makebox[0pt][r]{\@thefnmark\,}#1}
\makeatother 
\renewcommand{\figurename}{\small{Fig.}~}
\sectionfont{\large}
\subsectionfont{\normalsize} 

\fancyfoot{}
\fancyfoot[LO,RE]{\vspace{-7pt}\includegraphics[height=9pt]{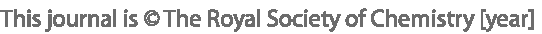}}
\fancyfoot[CO]{\vspace{-7.2pt}\hspace{12.2cm}\includegraphics{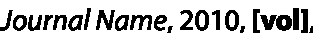}}
\fancyfoot[CE]{\vspace{-7.5pt}\hspace{-13.5cm}\includegraphics{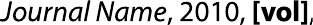}}
\fancyfoot[RO]{\footnotesize{\sffamily{1--\pageref{LastPage} ~\textbar  \hspace{2pt}\thepage}}}
\fancyfoot[LE]{\footnotesize{\sffamily{\thepage~\textbar\hspace{3.45cm} 1--\pageref{LastPage}}}}
\fancyhead{}
\renewcommand{\headrulewidth}{1pt} 
\renewcommand{\footrulewidth}{1pt}
\setlength{\arrayrulewidth}{1pt}
\setlength{\columnsep}{6.5mm}
\setlength\bibsep{1pt}

\twocolumn[
  \begin{@twocolumnfalse}
\noindent\LARGE{\textbf{First Principles Study of Photo-oxidation Degradation Mechanisms in P3HT 
  for Organic Solar Cells$^\dag$}}
\vspace{0.6cm}

\noindent\large{\textbf{Na Sai,\textit{$^{a,b}$} Kevin Leung,\textit{$^{c}$} Judit Z{\'a}dor,\textit{$^{d}$} and Graeme Henkelman\textit{$^{a}$}}}\vspace{0.5cm}

\noindent\textit{\small{\textbf{Received Xth XXXXXXXXXX 20XX, Accepted Xth XXXXXXXXX 20XX\newline
First published on the web Xth XXXXXXXXXX 200X}}}

\noindent \textbf{\small{DOI: 10.1039/b000000x}}
\vspace{0.6cm}

\noindent \normalsize{We present a theoretical study of degradation mechanisms for photoinduced oxidation in organic polymers in the condensed phase, using poly(3-hexylthiophene)(P3HT) as an example.  Applying density functional theory with a hybrid density functional and periodic boundary conditions that account for steric effects and permit  the modeling of interchain chemical reactions, we investigate reaction pathways that may lead to the oxidation of thiophene backbone as a critical step toward disrupting the polymer conjugation. We calculate energy barriers for reactions of the P3HT backbone with oxidizing agents including hydroxyl radical (OH$\cdot$), hydroperoxide (ROOH), and peroxyl radical (ROO$\cdot$),  following a UV-driven radical reaction starting at the $\alpha$-carbon of the alkyl side chain as suggested by infrared (IR) and X-ray photoemission (XPS) spectrosocopy studies. The results strongly suggest that an attack of OH$\cdot$ on sulfur in P3HT is unlikely to be thermodynamically favored. On the other hand, an attack of a peroxyl radical on the side chain on the P3HT backbone may provide low barrier reaction pathways to photodegradation of P3HT and other polymers with side chains.  The
condensed phase setting is found to qualitatively affect predictions of degradation processes.}
\vspace{0.5cm}
 \end{@twocolumnfalse}
  ]
\footnotetext{\textit{$^{a}$ Energy Frontier Research Center (EFRC:CST) and Institute for Computational Engineering and Sciences, The University of Texas at Austin, Austin, TX, 78712, USA; E-mail: nsai@physics.utexas.edu}}
\footnotetext{\textit{$^{b}$ Department of Physics, The University of Texas at Austin, Austin, TX 78712}}
\footnotetext{\textit{$^{c}$ Sandia National Laboratory, Albuquerque, NM 87185, USA }}
\footnotetext{\textit{$^{d}$ Sandia National Laboratory, Livermore, CA 94550 USA }}


\footnotetext{\dag~Electronic Supplementary Information (ESI) available. See DOI: 10.1039/b000000x/}

\section{Introduction}
Organic materials based on polymers
and small molecules have clear advantages such as light weight, mechanical flexibility, and
solution processibility, making them promising for large scale and low cost fabrication of next-generation electronic devices and solar cells.  In the past couple of years, organic photovoltaics (OPVs) have made rapid progress in power conversion efficiency (PCE) that has now surpassed
10\% in best performing research cells. With the promise of the synthesis of high performance
low-band gap polymers and fabrication of
multi-junction devices, it may be practical to expect the next PCE milestone (15\%) to be within reach in the next few years. On the other hand,  there remains a severe limitation in terms of the 
stability and lifetime of OPVs.\cite{Krebs:2012ty} In contrast to the intense research effort in improving PCE, fundamental understanding of the degradation mechanisms in OPV materials has not received much consideration. Only recently has the subject of stability and degradation mechanisms of OPV materials and devices started to gain more attention.\cite{Jorgensen:2008dt,Jorgensen:2011fk,Rosch:2012fg,Andreasen:2012cy,Brabec:2010hg,Grossiord:2012jk, Lee:2012uk,Peters:2011kd,Yip:2012it} Despite the importance of stability to OPV polymers, there has been little theoretical work on photochemical degradation mechanisms in OPVs.

Photo-oxidation is one of the
leading chemical degradation mechanisms in photoactive materials and occurs when organic materials are exposed to air and light. Chemical changes cause disruption of the $\pi$
conjugation of the polymers and reduce photoabsorbance in a process called photobleaching. Manceau {\it et al.} have ranked
the photochemical stability of a large number of molecular units that
are used in donor-acceptor copolymers and have found that the stability depends on the chemical 
structure.\cite{Manceau:2011km} 
However, some of the general rules identified, such as the critical role played by the side chains in the degradation process, seem to be valid for general OPV polymers.\cite{Manceau:2011km,Tournebize:2012cl}

\begin{figure*}
\centering
\includegraphics[width=15cm]{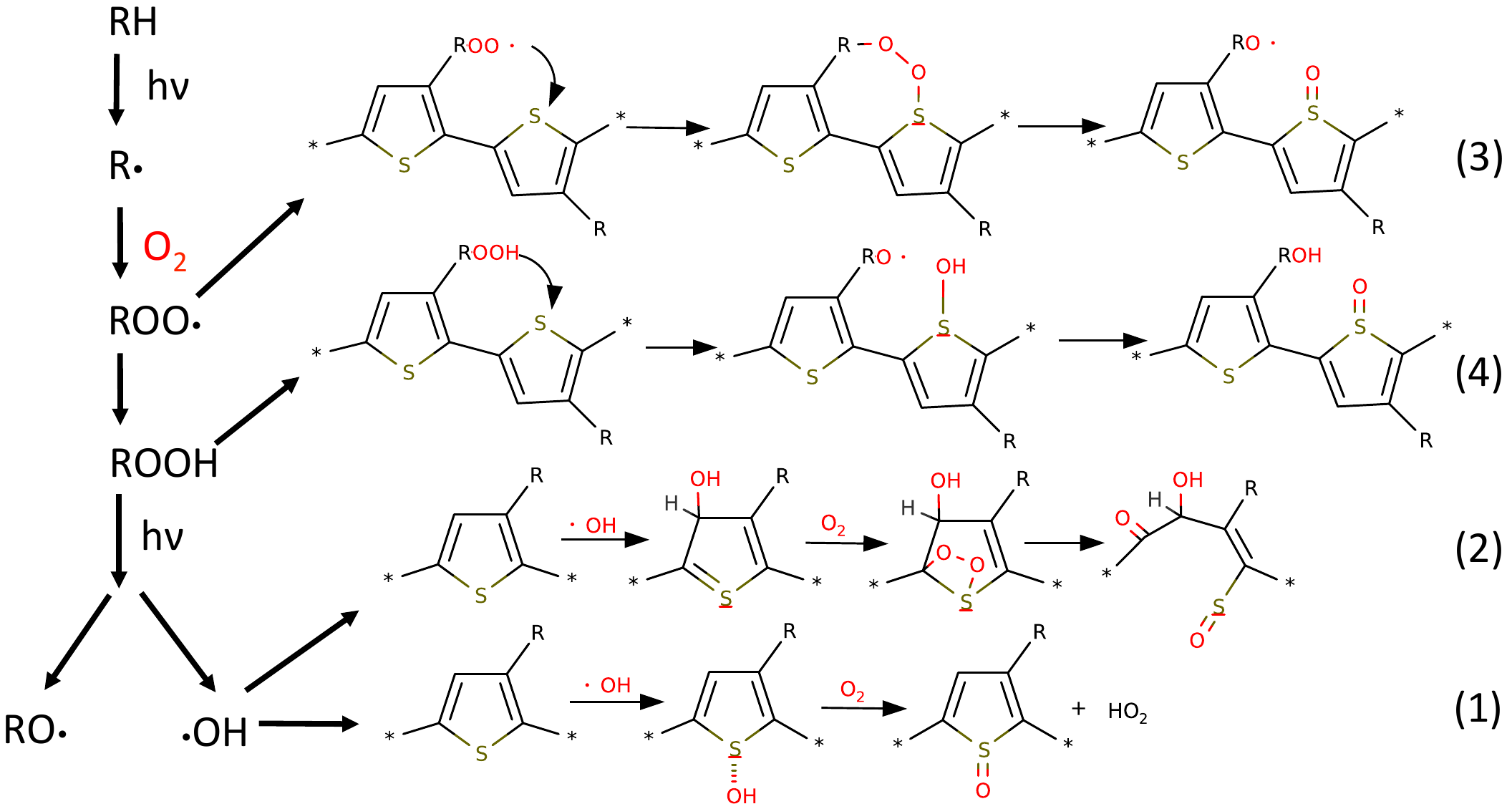}
\caption{Mechanisms for photo-oxidation reactions that are initiated on the alkyl side chain and subsequently lead to sulfur oxidation (S=O) on the backbone of the P3HT polymer. Mechanism (1) was suggested by Manceau {\it et al}.\cite{Manceau:2009kw} Our study shows that  mechanism (3) has the lowest reaction barrier and thus may potentially be the most relevant degradation pathway for P3HT among these mechanisms.}
\label{fig:mechanisms}
\end{figure*}

In this work, we choose poly(3-hexylthiophene)(P3HT) as our model system to study photo-oxidation induced degradation mechanisms in OPV polymers.  P3HT continues to serve as a prototypical benchmark system for understanding fundamental photovoltaic properties.  Both UV/visible and photoluminescence spectrosocopies have demonstrated photobleaching when P3HT films are irradiated in air.\cite{Manceau:2009kw,Hintz:2010eq,Hintz:2011ds,Deschler:2012ci} However, there remains controversy regarding which mechanisms are responsible for the photo-oxidation induced degradation.\cite{Abdou:1993vr,Abdou:1995uu,Abdou:1997ux,Manceau:2008kr, Reese:2008ey, Manceau:2009kw, Manceau:2010cb, Hintz:2010eq,Norrman:2010bq, Reese:2010ev,Hintz:2011ds,Hintz:2011kw,Sperlich:2011bl,Seemann:2011cd,Guerrero:2012dg,Hoke:2012ec,Hintz:2012vk,Cook:2012hw,Deschler:2012ci,Distler:2012wl,Tournebize:2013dy}  Earlier experiments on P3HT in solvents have
suggested photosensitization and reactions with singlet oxygen $^1$O$_2$ 
as the culprit.~\cite{Abdou:1993vr,Abdou:1995uu}  More recent investigations have shown evidence that the singlet is not the principal photo-oxidative intermediate of P3HT.\cite{Manceau:2008kr} Based on X-ray photoemission (XPS) and infrared-absorption
spectroscopy of P3HT solid films, a radical mechanism has been suggested. This
involves an initial oxidation on the alkyl side chain forming oxidative species such as hydroperoxide (ROOH), followed by
subsequent oxidation of thiophene that leads to the stepwise formation of sulfoxides $-$S$=$O, sulfones $-$SO$_2-$, or sulfinic acids $-$SO$-$OH.\cite{Norrman:2010bq, Manceau:2009kw,Manceau:2010cb,Hintz:2010eq,Hintz:2011ds,Tournebize:2012cl} 
These studies take the viewpoint that  critical to the overall degradation process is the initial formation of alkyl radicals R, and the reaction with molecular oxygen O$_2$ forming a variety of products including the alkylperoxyl radical (ROO),  alkyl hydroperoxide (ROOH), and hydroxyl radical (OH) that may at  a later stage become oxidizing agents responsible for the degradation reactions on the thiophene backbone (see Fig.~\ref{fig:mechanisms}). 
Furthermore, it has been suggested that polymers in an excited state may form a reversible charge transfer complex with O$_2$ and produce superoxide anions that can act as oxidizing radicals to attack the polymer.~\cite{Sperlich:2011bl,Seemann:2011cd,Guerrero:2012dg,Hoke:2012ec} 
Detailed understanding of the reaction mechanisms is important for calibrating the  structure-photochemical stability relationship and selecting and designing active materials with improved stability.  

In this study, we explore reactions between the side chain oxidation products and the thiophene backbone in P3HT using a combination of hybrid functional density functional theory (DFT) and quantum chemistry methods. 
Very few {\it ab-initio} studies of chemical reactions in polymer structures have been carried out in the condensed phase. As far as we know the only theoretical study of photo-induced defects in solid state OPV polymers was focused on the hydrogen addition defect with an activation barrier of 2 eV.\cite{Street:2012kr} 
Although the computation we present here does not guarantee a unique mechanism, especially because there may be more than one multistep reaction in solid state chemistry that contributes to photodegradation, we can interrogate
a few that have been proposed in the literature but not yet been directly investigated, and provide alternative suggestions that may enable new insights into polymer degradation chemistry.   In that sense, we follow an approach similar in spirit to the one applied in modeling the degradation of lithium ion battery electrolytes.\cite{e2}


Fig.~\ref{fig:mechanisms} illustrates the mechanisms we investigate in this work. In mechanism (1), we consider the reaction between a hydroxyl radical  and the thiophene in P3HT to form a SOH adduct. This is a common  reaction route for oxidation of sulfides such as dimethyl sulfide (DMS),\cite{Barnes:2006dl} the most abundant biogenic sulfur compound in the atmosphere; it has recently been proposed as a pathway for thiophene oxidation in P3HT.\cite{Manceau:2009kw}  In mechanism (2), we explore an alternative pathway of OH attack on thiophene on the carbon site to form a COH adduct. In the gas phase, the COH complex has been found to be significantly more stable than the SOH adduct in terms of thermodynamics.\cite{Barckholtz:2001wn} While a theoretical calculation of reactions between thiophene and triplet O$_2$  in gas phase  has found a barrier of 1.6 eV,\cite{Song:2012el} further reactions of COH adducts, including reaction with O$_2$ have never been reported. Here we present preliminary studies of the reactions in gas-phase models embedded in a continuum solvent field. In mechanisms (3) and (4), we employ a solid state model to explore the role of peroxyl radical and hydroperoxide on the side chain as oxidative agents for sulfur in P3HT. We show that explicit inclusion of the solid state polymer environment can lead to reactions not readily envisioned in traditional gas phase cluster-based calculations.

\section{Computational Model and Method}
All the solid state calculations were performed using DFT as implemented in the Vienna Ab-initio Simulation Package (VASP) with plane-wave basis set with a cutoff energy of 400 eV and projector augmented wave (PAW) potentials.\cite{Kresse:1996vf} Calculations were carried out with the Perdew-Burke-Ernzerhof (PBE)~\cite{Perdew:1996ug} and hybrid PBE0 functional with 25\% Hartree-Fock (HF) exchange.\cite{Perdew:1996tq,Adamo:1999hv} A hybrid functional with 50\% HF exchange was used in the calculation in Section 3.3 to examine the variation of barrier height as the fraction of exact exchange is changed. To search for minimum reaction energy paths, we applied the climbing-image nudged elastic band (NEB) method.\cite{Henkelman:2000ez} Further refinement of  the transition state is carried out using the dimer method.\cite{Henkelman:1999vr} To model the P3HT polymer solid, we adopt the crystalline structure of the regioregular head-to-tail poly(3-butylthiophene) (P3BT)\cite{Arosio:2008tw} (orthorhombic $C222_1$ space group with lattice parameters $a$ = 7.64 \AA, $b$ = 7.75\AA, $c$ = 24.97\AA, $\beta = 90.0^\circ$). Unless specified, the model contains a periodically repeated P3BT supercell that doubles the unit cell along the backbone ($b$-axis) direction with 16 thiophene rings and 306 atoms. A $2\times1\times1$ and  $2\times2\times1$ Monkhorst-Pack $k$-point sampling was used for the P3BT supercell and unit cell, respectively. 

Geometry relaxations for isolated molecules and molecular complexes were performed in Gaussian 09~\cite{g09} with DFT/PBE, DFT/PBE0 (with or without the D3(BJ) dispersion correction),~\cite{Grimme:2011cl} and the second-order M{\o}ller-Plesset (MP2) method, all using the aug-cc-pvdz basis set. Single point calculations at the coupled cluster CCSD(T)/aug-cc-pvdz level of theory were applied to check the accuracies of DFT. The Gibbs free energy formation is obtained by taking into account the zero-point energy correction and thermal corrections after performing frequency and normal mode analysis. A dielectric solvent with $\epsilon=3$ treated in a continuous surface charge polarizable continuum model (CSC-PCM) was applied unless otherwise noted.~\cite{Scalmani:2010el} 

In order to justify the relevance of a specific reaction to the degradation of OPV polymers, we have devised a cutoff for the activation energy such that if the theoretically calculated barrier is much higher than the cutoff, we abandon the reaction mechanism. Applying the Arrhenius relation for the reaction rate $k = A \exp(-E_{\rm A}/k_BT)$ and assuming an operation temperature at 298 K, a reaction prefactor of $10^{12} s^{-1}$ (Eyring constant), and an OPV life time of 100 hours, we estimate the upper bound for the activation energy $E_{\rm A}$ should be $\le$ 1 eV for the reaction to be relevant to photo-oxidation degradation in OPV polymers.  This criterion assumes all reactants are in thermal equilibrium.  It ignores the fact that very fast reactions can
be initiated by energetic, photo-absorbing species such as the photo-generated OH radical.  Theoretical studies for generating such initial
products are the domain of time-dependent DFT (TDDFT).  In that sense, our
work focuses on the later-stage, slower reactions where such initial products have
had time to equilibrate.

\section{Results and discussion}

\subsection {Thiophene Reactions with OH in P3HT}
Extensive experimental and {\it ab-initio} theoretical studies have established that the addition of OH  is the initial step for the oxidation of sulfides such as DMS.\cite{Barnes:2006dl} The reaction goes through an initial formation of a DMS$-$OH complex with a S$-$O bond with a bond strength ranging from 6 to 9 kcal/mol. The sulfur oxidation reaction then proceeds through a reaction with O$_2$ (i.e., \ce{DMSOH + O$_2$ $\rightarrow$ DMSO + HO$_2$}).  It has been suggested that a similar reaction route as in mechanism (1) of Fig.~\ref{fig:mechanisms} may explain the sulfur oxidation in photodegradation in P3HT and the S$=$O signature observed in XPS.~\cite{Manceau:2009kw} Here we apply DFT in combination with quantum chemical calculations to examine the thermodynamics of this reaction route. 

\begin{figure}
\includegraphics[width=8.5cm]{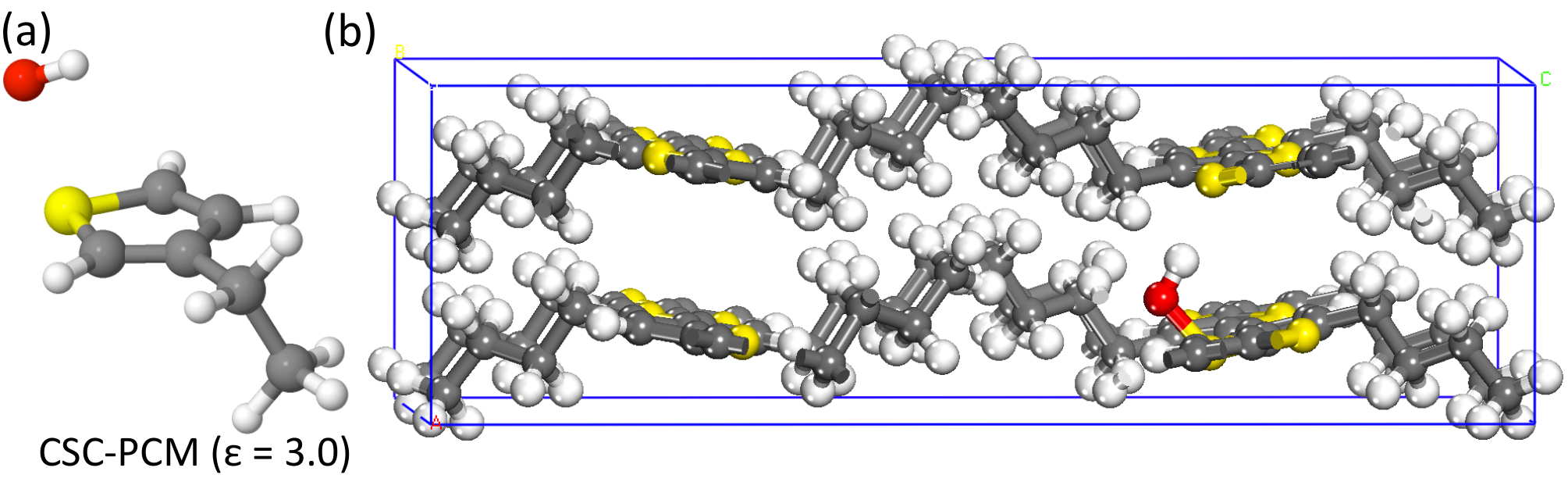}
\caption{Schematic of the P3HT-OH complex as in mechanism (1), represented by (a) gas phase ethylthiophene$-$OH embedded in a continuum solvent described by a contious surface charge polarizable contniuum model (CSC-PCM) with a dielectric constant $\epsilon=3.0$ and (b) periodic super cell of poly(3-butylthiophene).}
\label{fig:SOH}
\end{figure}
\begin{table}[h]
\small
  \caption{Binding energies (eV) and bond distances (\AA) for ethylthiophene-OH embedded in a dielectric solvent (Fig.~\ref{fig:SOH}a), a model system used in mechanism (1), calculated using PBE and PBE0/aug-cc-pvdz in Gaussian 09 (``$-$D3BJ''  includes Grimme's dispersion correction). $E$//PBE0 denotes binding energy using single point energy at the fixed PBE0 geometry.  $\Delta G$ is the free energy of formation at $298.15$ K, obtained as discussed in the text.}
  \label{tbl:TOH-DFT}
  \begin{tabular*}{0.5\textwidth}{@{\extracolsep{\fill}}lllll}
    \hline
    Method &$d_{\rm S-O}$  & $d_{\rm S-H}$  & $E_{\rm b}$ & $\Delta G $  \\
    \hline
    PBE & 2.19 & 2.49& $-$0.56 & \\
    PBE$-$D3BJ & 2.20 & 2.51& $-$0.62 & \\
    PBE0 & 2.32 & 2.57 & $-$0.22 & 0.05\\
  PBE0$-$D3BJ & 2.32 & 2.56 & $-$0.27 & \\
    MP2//PBE0 & & & $\phantom{-}$0.24 &\\
    MP4//PBE0& & & $\phantom{-}$0.14 & \\
    CCSD(T)//PBE0 & & & $-$0.07 & \\
    \hline
 \end{tabular*}
\end{table}
In out first step toward validating our modeling of SOH formation in  P3HT, we carried out a study using ethylthiophene-OH, as shown in Fig.~\ref{fig:SOH}a, embedded in a solvent reaction field. Table~\ref{tbl:TOH-DFT} shows the optimized S$-$O and S$-$H bond distances and electronic binding energies $E_b$ calculated using PBE and PBE0. The  ``DFT-D3BJ'' functional which includes Grimme's dispersion correction~\cite{Grimme:2011cl} is also used. Using PBE0, we tested the basis set convergence by performing calculations with aug-cc-pvdz, aug-cc-pvtz, and aug-cc-pvqz basis sets. We found that the S$-$O bond length sightly reduces while the binding energy already converges at aug-cc-pvdz with an extrapolated value of $-$0.21 eV at the basis set limit (see Table S1). In comparison, PBE considerably overestimates the binding energy.  The dispersion correction does not affect the geometries except for slightly strengthening the binding energies. Henceforth we will omit dispersion correction in all calculations. Using single point energy calculations at fixed PBE0 geometry (all with aug-cc-pvdz), we verified that the SOH complex becomes either unbound in MP2 and MP4 or weakly bound with a binding energy of $-$0.07 eV in CCSD(T). The latter value is numerically comparable to the PBE0 value within a quantum-chemistry accuracy, suggesting that PBE0 is reasonably reliable in describing the weakly bound complexes. 

Interestingly, geometry optimization using unrestricted HF and MP2/aug-cc-pvdz predicts that the H atom of OH is closer to sulfur (see Fig. S1). This is reminescent of a local minimum previously reported for DMS$-$OH in MP2 calculations that did not converge on basis sets.\cite{Gross:2004wx} Such a minimum could not be stabilized with any DFT calculations  including those that incorporated dispersion corrections. However since HF/aug-cc-pvtz also yields the “H-down” geometry,  we suggest that the H-down geometry may be a metastable minimum for ethylthiophene-OH that is favored by HF. Using CCSD(T) single point energies at fixed MP2 geometry, we found a binding energy of $-$0.18 eV  (Table S2) which is similarly small as the CCSD(T) binding energy at the PBE0 geometry, suggesting that ethylthiophene-OH is a weakly bound complex regardless of geometry. 

Applying a periodic solid state model for P3HT based on the crystalline poly(3-butylthiophene) as shown in Fig.~\ref{fig:SOH}b, we calculated the binding energies and bond distances for the P3HT-OH complex using PBE and PBE0 in VASP (Table~\ref{tbl:OHbinding}). Compared to PBE0, PBE overestimates the binding energy, similar as in the Gaussian 09 calculations. Moreover, we find that SOH (with a binding energy of $-$0.24 eV with PBE and being unbound with PBE0) is significantly less bound in the solid than in the gas phase. We attribute the difference between the gas phase  and the solid to a steric hindrance effect associated with the bulky side chains in the solid state which should be included in modeling the reactivity of polymers.   

\begin{table}
\caption{Binding energies (eV) and bond distances (\AA) for the P3HT-OH complex (Fig.~\ref{fig:SOH}b) calculated using periodic supercell and PBE and PBE0.  $\Delta G$ is calculated for PBE0 only.}
\label{tbl:OHbinding}
  \begin{tabular*}{0.5\textwidth}{@{\extracolsep{\fill}}lllll} 
    \hline
 Method  &$d_{\rm S-O}$  & $d_{\rm S-H}$  & $E_{\rm b}$ & $\Delta G$ \\
    \hline
     PBE & 1.90 & 2.33 & $-$0.24 & \\
    PBE0  & 1.66 & 2.19 & $\phantom{-}$0.07 & 0.34 \\
    \hline
 \end{tabular*}
\end{table}
Finally, we calculated the entropy correction by carrying out  frequency analysis. Using gas phase ethylthiophene-OH, we found a thermal energy correction of 0.4 eV at 298 K and 1.0 atm, coming mostly from a loss of translational entropy in the bimolecular reaction. The entropy correction for the solid is calculated by accounting the difference between the concentration of sulfur in the ideal gas molecule and the solid (molar volume V$_{\rm gas}$  = 24.465 L/mol and V$_{\rm P3HT}$ =0.15 L/mol) and estimated to be 0.27 eV. Applying this correction to P3HT-OH, we find a free energy of formation of 0.34 eV.  Thus the SOH formation reaction in P3HT is endothermic.

\subsection{Alternative Reaction Channels with OH in P3HT}
\label{sec:coh}
\begin{figure}
\includegraphics[width=8.5cm]{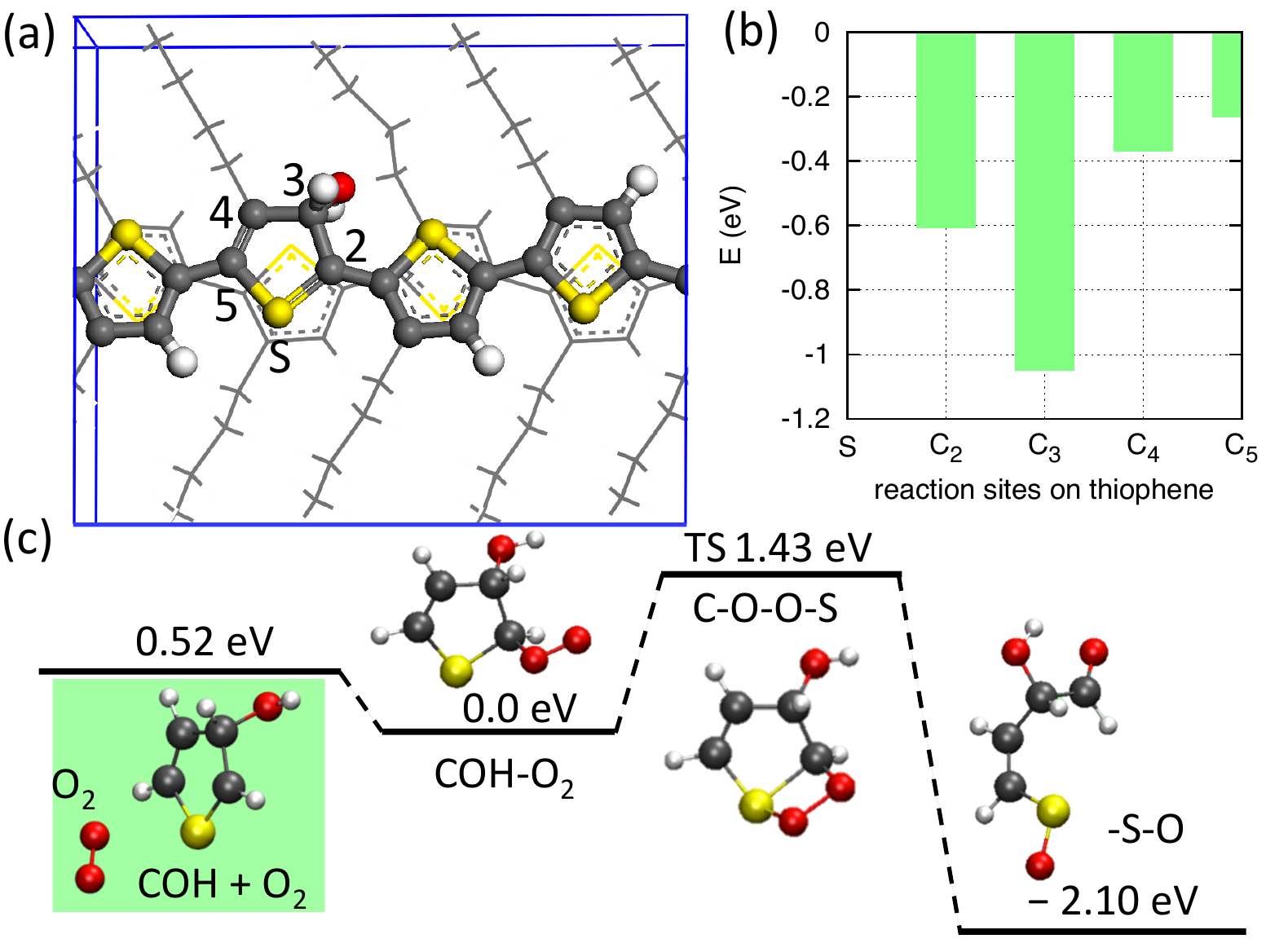}
\caption{(a) The most stable C$^{(3)}$OH complex in P3HT  (only half of the  supercell in which OH resides is shown). (b) Comparison of the reaction energy of the P3HT-OH complex on the sulfur (S) and carbon (C) sites of the thiophene ring as determined by PBE0 in VASP-PAW calculations. The energy for SOH has been zeroed out. (c) Reaction pathway of an O$_2$ triplet attack on a C$^{(3)}$OH complex as shown in mechanism (2). All energies include a dielectric solvent of $\epsilon = 3$ and entropy correction.}
\label{fig:P3BTOH}
\end{figure}

In this section we investigate alternative OH reaction channels in P3HT as illustrated in mechnism (2). Using the PBE0 functional and the P3BT supercell, we calculate the formation energy of COH, the product of the reaction between OH and the carbon sites in the thiophene ring. Fig.~\ref{fig:P3BTOH}b shows comparison between the binding energies of C$^{(i)}$OH, where $i=2,3,4,5$, and SOH. The most stable complex  is C$^{(3)}$OH (see Fig.~\ref{fig:P3BTOH}a) with a formation energy lower than that of SOH by more than 1 eV. This energy landscape is somewhat different from gas phase~\cite{Barckholtz:2001wn} where the most stable OH complex is on C$^{(2)}$, the carbon immediately next to sulfur. This may be attributed to a steric hindrance effect where the presence of conjugation along the thiophene backbone  in P3HT  has restricted the reactivity of C$^{(2)}$.  

We can state the following about the OH complex in P3HT:

(1.)\ The  attack of OH  on sulfur in P3HT is endothermic and  the P3HT(S)-OH complex is thermodynamically unstable.

(2.)\ Alternative OH reaction channels through the nearby carbon sites on the thiophene ring are much more favorable than those involving SOH, with the most stable COH complex lying more than 1 eV below SOH.  

(3.) The SOH can relax into the COH configuration thermodynamically  without overcoming a significant activation barrier  (see Supporting Information). Therefore even if SOH forms, perhaps as a transient during the reaction process, it will fall into the potential well of the COH configuration.    

Thus unlike in DMS, formation of the SOH complex as suggested by the recent experimental work,~\cite{Manceau:2009kw} is unlikely to be an energetically favored intermediate step for the subsequent oxidation reaction in P3HT. This emphasizes the pitfall of using chemical intuition based on an another
small molecule to predict the degradation pathway of P3HT.

To examine whether the COH complex may serve as an intermediate for subsequent reactions with O$_2$, we  carried out a preliminary study of the most stable COH complex reacting with O$_2$ using gas phase models embedded in a dielectric solvent of $\epsilon=3$ (Fig~\ref{fig:P3BTOH}c). Following a recent theoretical study of reaction pathways of an O$_2$ attack on thiophene by Song {\it et al.},\cite{Song:2012el} we investigated a $2+2$ side-on addition pathway in which the two oxygen atoms of an O$_2$  triplet add to the C$^{(2)}-$S bond of C$^{(3)}$OH. This is the only reaction pathway that  leads to sulfur oxidation. Using PBE0/6-31G(d,p), we calculated the energy profiles of the reaction (see Fig~\ref{fig:P3BTOH}c). The formation of a linear sulfine product is highly exothermic. Compared to the bare thiophene,\cite{Song:2012el}  the reaction with the O$_2$ triplet has a lower reaction barrier, possibly due to the doublet spin configuration of the COH.  However the reaction barrier (1.43 eV) is still rather high, making this reaction unlikely based on our criterion for  reactions relevant to P3HT degradation. We also investigated a $2+4$ cycloaddition reaction in which the two O atoms add to C$^{(2)}$ and C$^{(5)}$ and found a barrier of 1.1 eV (see Fig. S3), but it does not produce sulfur oxidation. 

In view of the unsuccessful search for low barrier reactions associated with the OH radicals that yield the S$=$O motif, we must turn to alternative degradation routes.

\subsection{Alkylperoxyl (${\rm ROO}\cdot$) + Thiophene Reaction }
\label{sec:ROO}

It has been suggested that the initial photo-oxidation processes in P3HT takes place on the  side chain~\cite{Manceau:2010cb,Manceau:2009kw,Hintz:2010eq} and leads to reactive species that may potentially oxidize the polymer backbone. Although these initial processes have not been theoretically investigated for P3HT, the reactions are likely initiated by a radical formation on the alkyl chains (\ce{RH ->[\ce{h\nu}] R^{.} + H^{.}}).\cite{Rabek:1996jx} This step requires a UV-driven photo-dissociation of a side-chain C$-$H bond at the $\alpha$-position (the carbon atom directly bonded to the thiophene ring), known to be more reactive as a result of resonance stabilization. This is then followed by chain propagation steps involving (1)  ${\rm R}\cdot \chemsign{+}{\rm O}_2 \rightarrow \rm{ROO}\cdot$  in which the alkyl radical  reacts with O$_2$ to form the alkylperoxyl radical  and (2) ${\rm ROO}\cdot\chemsign{+} {\rm RH} \rightarrow {\rm ROOH} \chemsign{+} {\rm R}\cdot$ in which the peroxyl radical proceeds to form a hydroperoxide by abstracting an H atom  from the same or a nearby alkyl chain. 

\begin{figure}[h]
\includegraphics[width=8.5cm]{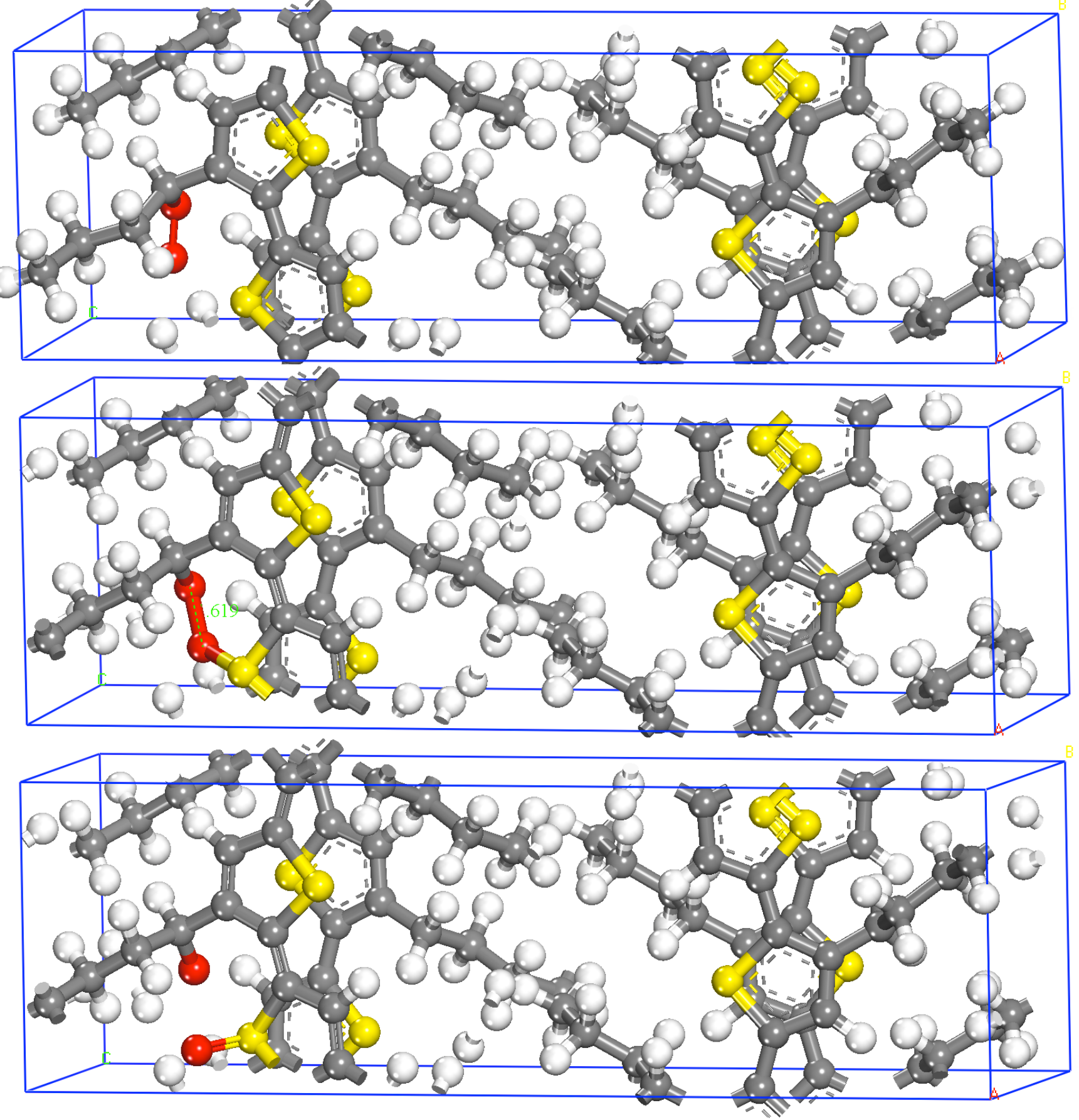}
  \caption{The P3BT unit cell in which the alkylperoxyl radical ${\rm ROO}\cdot$ on the side chain attacks the sulfur atom on a nearby thiophene ring as in mechanism (3). The three panels show the initial reactants, the transition state in which the O$-$O bond breaks, and the formation of sulfoxide (S=O) and alkoxy ${\rm RO}\cdot$ on the side chain.}
 \label{fig:cell_PBE}
\end{figure}

In this section, we investigate the reaction pathway 
\begin{equation}
\setatomsep{3em}
\setchemrel{0pt}{0.5em}{2em}
\lewis{0.,ROO}  \chemsign{+} \chemfig{S*5([,0.5]-=-=-)} \chemrel{->} \chemfig{S*5([,0.5](=O-RO)-=-=-)}  \chemrel{->}   \lewis{0.,RO} \chemsign{+} \chemfig{S*5([,0.5](=O)-=-=-)}  
\label{eq:ROO}
\end{equation}
in which the alkylperoxyl radical is a reactant for sulfoxide formation in P3HT as in mechanism (3). As Eq.~\ref{eq:ROO} shows, this may occur through the formation of a transition state forming a RO$-$O$=$S bond. In the final product, the O$-$O bond breaks and leads to the formation of alkoxy ${\rm RO}\cdot$ and sulfoxide S$=$O as shown in Fig.\ref{fig:cell_PBE}. 
First let us show that ${\rm ROO}\cdot$  is thermodynamically stable  in P3HT and the competing reaction 
\begin{equation}
\setatomsep{3em}
\setchemrel{0pt}{0.5em}{2em}
\chemfig{\lewis{0.,ROO}  \chemsign{+} RH \chemrel{->} ROOH \chemsign{+}\lewis{0.,R}}
\label{eq:RH}
\end{equation}
that depletes ${\rm ROO}\cdot$ is less favorable than Eq.~(\ref{eq:ROO}).

Using a P3BT supercell that doubles the unit cell along the backbone direction and PBE0, we find that the  ${\rm R}\cdot\chemsign{+}{\rm O}_2 \rightarrow\rm{ROO}\cdot$ formation is exothermic with a reaction energy of $-$0.67 eV. This is a shallower potential well than the $-$1.43 eV minimum in the gas phase~\cite{Zador:2011kz} but is sufficient to ensure that ${\rm ROO}\cdot$ exists as a stable species in equilibrium with ${\rm R}\cdot$ and O$_2$. The competing reaction in Eq.~(\ref{eq:RH}), on the other hand,  is significantly endothermic with positive reaction  energies of 0.8 eV and 1.0 eV (similar to the widely accepted values in analogous gas-phase systems)~\cite{Zador:2011kz} for abstracting an H atom from the same and nearby alkyl chain, respectively. Indeed, production of ${\rm ROO}\cdot$ has been noted to happen much faster than the H abstraction reaction.\cite{Audouin:1994tu} Other pathways for depleting the ${\rm ROO}\cdot$ were also shown to be endothermic.\cite{Zador:2011kz} In contrast, the reaction in Eq.~(\ref{eq:ROO}) is exothermic with a reaction energy of $-0.53$ eV. Thus it may be the most important reaction involving ${\rm ROO}\cdot$ in P3HT. 

\begin{figure}
\includegraphics[width=8.5cm]{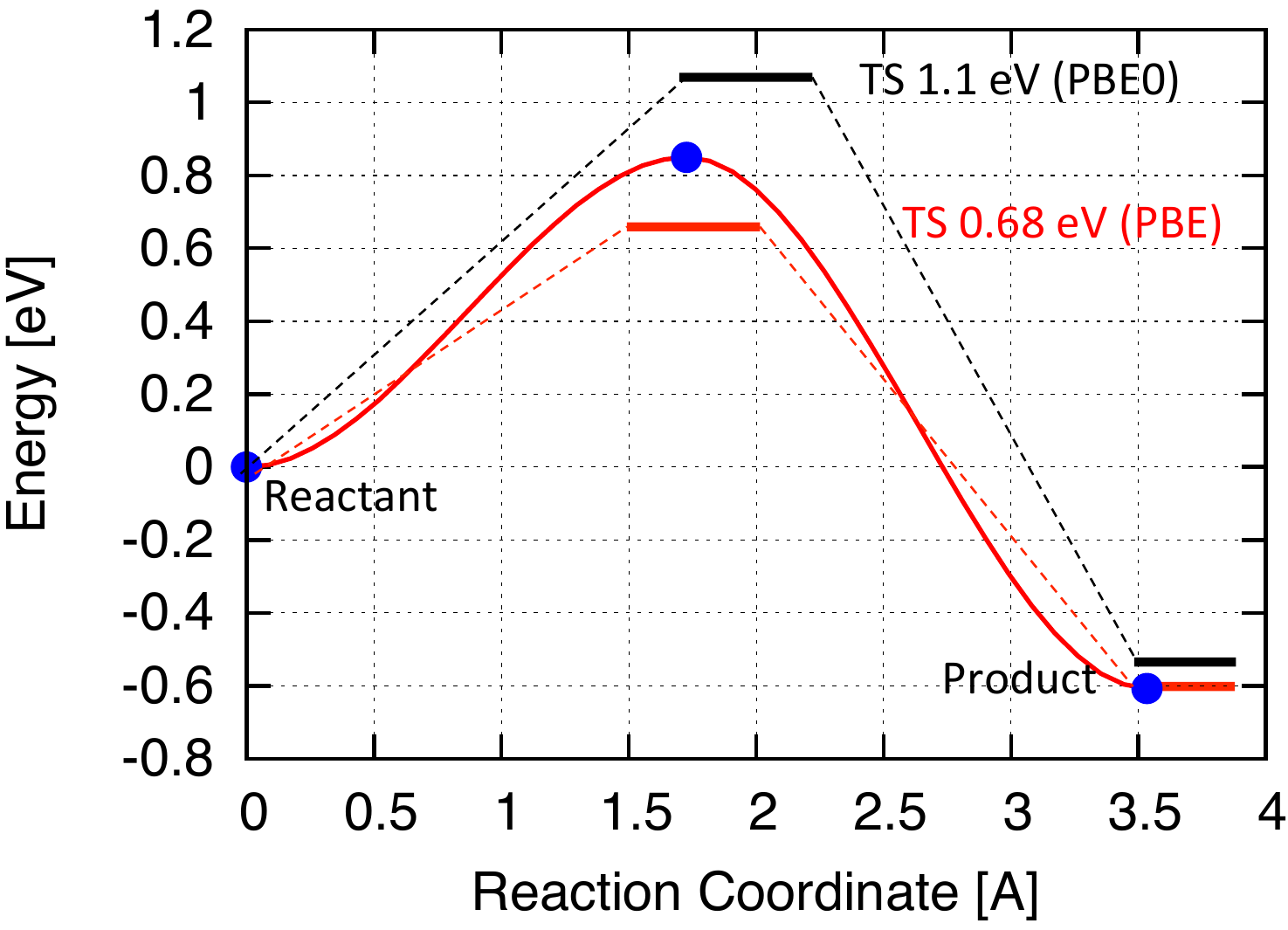}
\caption{Energy profile for mechanism (3) in which an alkylperoxyl (ROO) radical  attacks thiophene in the P3BT unit cell. Red solid line shows the minimum energy path obtained with NEB and PBE. The saddle points refined with the dimer method correspond to reaction barriers of 0.68 eV in PBE and 1.1 eV in PBE0. }
\label{fig:barrier}
\end{figure}

Fig.~\ref{fig:barrier} shows the reaction energy profile along the reaction coordinates calculated using the nudged elastic bands (NEB) method and P3BT unit cell (see  Fig.~\ref{fig:cell_PBE}). The NEB method performs a globally constrained search for  energy extrema along an initially linearly interpolated path between the reactant and product. To refine the transition state, we also applied the dimer method which does a local search by optimizing a vector initially connecting two local images near the saddle point. Using PBE, we found a barrier height of 0.68 eV (dimer method)  with an optimized O$-$O bond length of 1.61 \AA. Standard density functionals such as PBE are known to underestimate the reaction barriers by $4-8$ kcal/mol.\cite{Cohen:2012fm} The error can be significantly reduced  by applying hybrid functionals which mix a fraction of nonlocal HF exchange with semi-local exchange to partially cancel the delocalization error.\cite{Cohen:2008wh}  Applying the PBE0 hybrid functional and the dimer method we find a barrier height of 1.1 eV corresponding to a transition state with a O$-$O bond length of 1.7\AA. 

As the transition state search is computationally demanding, we limit the hybrid functional calculations to the P3BT unit cell as shown in Fig~\ref{fig:cell_PBE}. We considered  the effect of the simulation cell size by comparing the reaction barrier for a unit cell with that of a supercell that doubles the unit cell along the backbone direction within PBE calculations (see Fig. S4). The reaction barrier shifts down by 30\% from 0.85 eV  to 0.6 eV as a result of the strain relaxation afforded by the longer backbone chains in the supercell.  Zero point energy corrections, not included in these calculations, should also slightly reduce the transition state energy.  We therefore conclude that the activation barrier for the ROO reaction in Eq.~(\ref{eq:ROO}) should drop to about 1 eV or below when using the PBE0 functional and even PBEh with 50\% exchange,~\cite{note} making this reaction potentially relevant in the photo-oxidation of P3HT.  

\subsection{Alkyl hydroperoxide (ROOH) + Thiophene Reaction}
Finally we examined the reaction pathway in which hydroperoxide (ROOH) on the alkyl side chain reacts with thiophene in P3HT as mechanism (4). Although ROOH has been identified as a  photo-product in P3HT,\cite{Manceau:2009kw, Hintz:2010eq} the pathway for ROOH formation remains unclear. A pathway like Eq.~(\ref{eq:RH}) may be assisted by vibrational excitation in gas phase, but seems difficult in the solid given the endothermicity of the reaction. We speculate that a water (H$_2$O) contaminant in  P3HT samples may have contributed to the ROOH formation by reacting with the ROO radical. 

The proposed reaction pathway between ROOH and thiophene in P3HT is 
\begin{equation}
\setatomsep{3em}
\setchemrel{0pt}{0.2em}{1.5em}
\chemfig{ROOH}\chemsign{+}\chemfig{S*5([,0.5]-=-=-)}\chemrel{->}\lewis{0.,RO}\chemsign{+}\chemfig[scale=0.75]{S*5([,0.5](-\lewis{0.,HO})-=-=-)}\chemrel{->}\chemfig{ROH}\chemsign{+}\chemfig{S*5([,0.5](=O)-=-=-)} 
\label{eq:ROOH}
\end{equation}
in which ROOH attacks sulfur to form an intermediate SOH product and a RO radical on the side chain.  This is followed by proton transfer from the SOH group to the RO radical to result in a sulfoxide (S=O) on the backbone and a ROH (alcohol) group on the side chain that have been identified by the recent IR studies in P3HT.\cite{Manceau:2010cb,Manceau:2009kw} 
\begin{figure}[t]
\includegraphics[width=8.5cm]{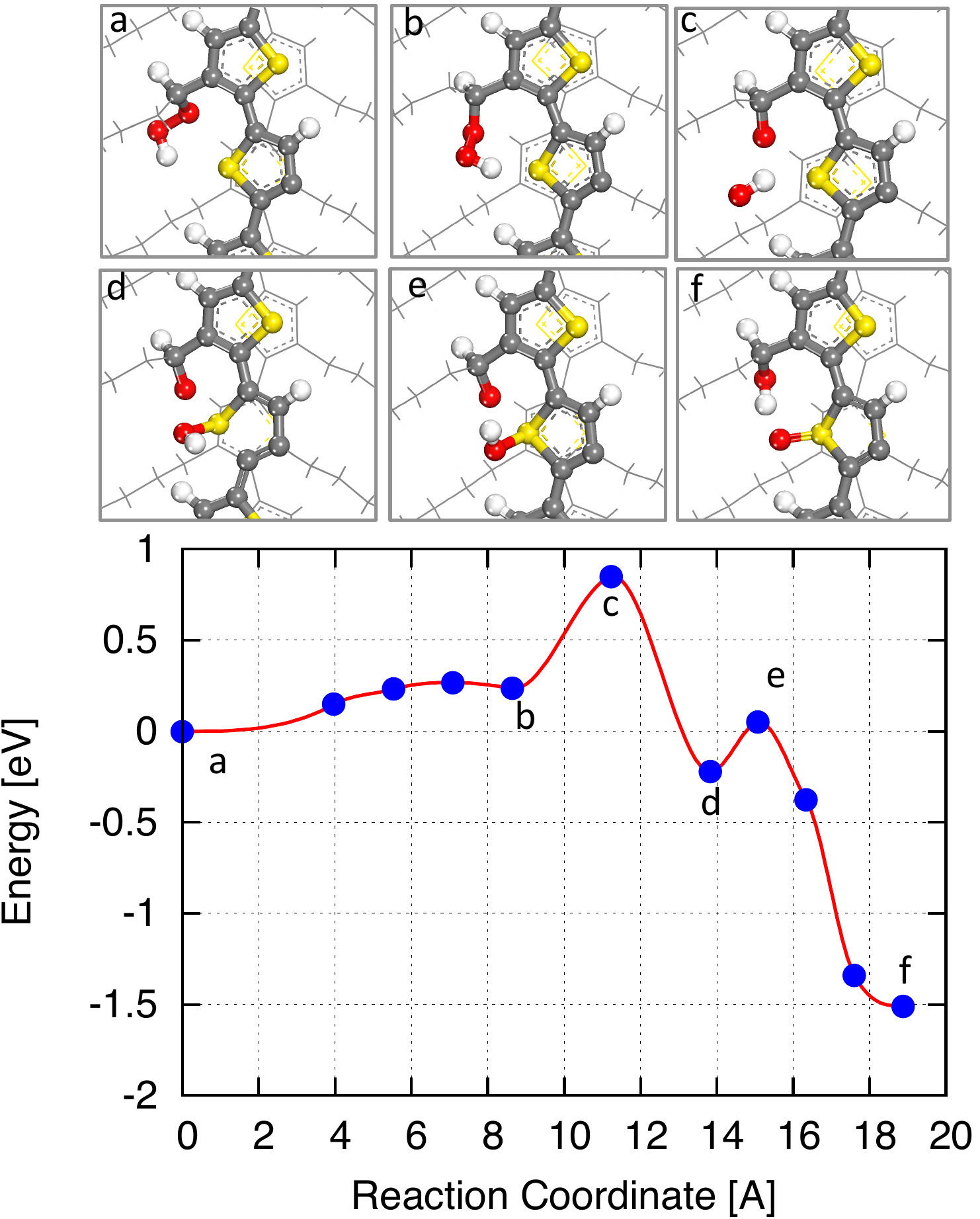}
\caption{ Minimum energy path for mechanism (4) calculated using a P3BT supercell doubled along the backbone direction and PBE. (a-f), respectively correspond to the reactant, the intermediate step after rotation of the ROOH bond, the bond breaking of ROOH, the intermediate step involving RO and SOH,  proton transfer, and the final product with S=O on thiophene and ROH on the side chain.}
\label{fig:ROOHbarrier}
\end{figure}

Fig.~\ref{fig:ROOHbarrier} illustrates the minimum energy path calculated using PBE and the NEB method in a P3BT supercell doubled along the backbone direction. The reaction turns out to be rather exothermic with a reaction energy of $-$1.5 eV. Following the potential energy profile, we have identified three reaction barriers for the process, including a rotation of the ROOH group on the alkyl side chain, a bond breaking of the RO$-$OH  group to form a SOH complex on the thiophene and a RO radical on the alkyl chain, and a proton transfer from the SOH complex to the RO group. The rotation and proton transfer both involve a small barrier of 0.25 eV, but breaking the RO$-$OH bond requires overcoming a reaction barrier of 0.85 eV. This is significantly greater than the barrier (0.6 eV) for the reaction in Eq.~(\ref{eq:ROO}).  Although we have not performed saddle point search on this reaction with PBE0 owing to the large size of the supercell and the cost of the calculations, we estimate that the barrier with PBE0 should shift up to at least 1.35 eV  based on our previous comparison of PBE and PBE0 in Sec. \ref{sec:ROO}. The ROOH reaction in Eq.~(\ref{eq:ROOH}) is thus going to be far less relevant compared to the  ${\rm ROO}\cdot$ mechanism  as a degradation pathway for P3HT. 

\section{Conclusions}
Using periodic solid state structures and DFT with the PBE0 hybrid density functional, we have performed a theoretical investigation of photoinduced oxidation reaction mechanisms that may be potentially relevant to photodegradation in P3HT OPV polymers. Following an initial photo induced  oxidation on the alkyl side chain that forms various  oxidizing groups such as the alkylperoxyl (${\rm ROO}\cdot$), alkyl hydroperoxide (ROOH), and hydroxyl (${\rm OH}\cdot$) radicals, we have investigated the reactions between thiophene and these oxidizing groups as pathways for the oxidation of the thiophene backbone (i.e., formation of S$=$O)  in P3HT.  

The calculated free energy of SOH adduct in P3HT shows that OH addition to the sulfur group is thermodynamically disfavored. In addition, the presence of much more favorable OH reaction channels on the carbon sites (through formation of COH adducts) and the negligible barrier between the SOH and COH complexes suggest that, unlike in the sulfides such as DMS,  the OH addition to sulfur is unlikely the preferred reaction  mechanism for the oxidation of thiophene backbone. We also investigated  reactions between the COH adducts and O$_2$ in clusters embedded in a continuum solvent ($\epsilon=3$), but were not able to identify an OH-led pathway leading to a low barrier sulfoxide (S$=$O) formation. 

We investigated alternative reactions involving the peroxyl radical (ROO) and hydroperoxide (ROOH) on the alkyl side chain attacking sulfur in thiophene in P3HT. The reaction energies and barrier heights calculated within the hybrid functional and transition state theory show that both ROO and ROOH may react exothermically with thiophene and form sulfoxide. However the barrier height for the ROO reaction route ($\lesssim1$ eV) is considerably lower than that for the ROOH route, making the ROO route the most preferred among the mechanisms we have considered for P3HT degradation under equilibrium conditions at 298K.  
Our theoretical results further corroborate the critical role of the side chain in polymer degradation processes that has been reported by the experiments.\cite{Manceau:2011km}
 
Even though hybrid functional DFT calculations are costly in the 
periodically replicated simulation cells adopted herein, we show that they are viable, and are particularly valuable in examining reactions between the backbone and the functional groups/radicals on side chains in close proximity.  Such reactions may have been overlooked
in the past simply because gas phase cluster-based theoretical models
examined in the literature lack the spatial correlation between such reacting species.  
Given how few theoretical calculations of chemical reactions in polymer materials have been carried out in the solid state, the work reported here demonstrates the potential of   theoretical studies of previously unexamined reaction mechanisms
responsible for photochemical degradation of novel light harvesting materials.

\section*{Acknowledgements}
NS thanks Christopher W.  Bielawski, Bradley J. Holliday, and Peter J. Rossky for valuable conversations. This work was supported as part of the Understanding Charge Separation and Transfer at Interfaces in Energy Materials (EFRC:CST), an Energy Frontier Research Center funded by the US Department of Energy, Office of Science, Office of Basic Energy Sciences under Award Number DE-SC0001091. Computing resources were provided by the National Energy Research Scientific Computing Center (NERSC) and the Texas Advanced Computing Center (TACC).  J. Z. is supported by the Division of Chemical Sciences, Geosciences, and Biosciences, the Office of Basic Energy Sciences, the U.S. Department of Energy. Sandia is a multiprogram laboratory operated by Sandia Corporation, a Lockheed Martin Company, for the National Nuclear Security Administration under contract DEAC04-94-AL85000. 




\footnotesize{
\bibliography{rsc} 

\providecommand*{\mcitethebibliography}{\thebibliography}
\csname @ifundefined\endcsname{endmcitethebibliography}
{\let\endmcitethebibliography\endthebibliography}{}
\begin{mcitethebibliography}{56}
\providecommand*{\natexlab}[1]{#1}
\providecommand*{\mciteSetBstSublistMode}[1]{}
\providecommand*{\mciteSetBstMaxWidthForm}[2]{}
\providecommand*{\mciteBstWouldAddEndPuncttrue}
  {\def\EndOfBibitem{\unskip.}}
\providecommand*{\mciteBstWouldAddEndPunctfalse}
  {\let\EndOfBibitem\relax}
\providecommand*{\mciteSetBstMidEndSepPunct}[3]{}
\providecommand*{\mciteSetBstSublistLabelBeginEnd}[3]{}
\providecommand*{\EndOfBibitem}{}
\mciteSetBstSublistMode{f}
\mciteSetBstMaxWidthForm{subitem}
{(\emph{\alph{mcitesubitemcount}})}
\mciteSetBstSublistLabelBeginEnd{\mcitemaxwidthsubitemform\space}
{\relax}{\relax}

\bibitem[Krebs(2012)]{Krebs:2012ty}
F.~C. Krebs, \emph{{Stability and Degradation of Organic and Polymer Solar
  Cells}}, John Wiley {\&} Sons, 2012\relax
\mciteBstWouldAddEndPuncttrue
\mciteSetBstMidEndSepPunct{\mcitedefaultmidpunct}
{\mcitedefaultendpunct}{\mcitedefaultseppunct}\relax
\EndOfBibitem
\bibitem[Jorgensen \emph{et~al.}(2008)Jorgensen, Norrman, and
  Krebs]{Jorgensen:2008dt}
M.~Jorgensen, K.~Norrman and F.~C. Krebs, \emph{Solar Energy Materials and
  Solar Cells}, 2008, \textbf{92}, 686--714\relax
\mciteBstWouldAddEndPuncttrue
\mciteSetBstMidEndSepPunct{\mcitedefaultmidpunct}
{\mcitedefaultendpunct}{\mcitedefaultseppunct}\relax
\EndOfBibitem
\bibitem[Jorgensen \emph{et~al.}(2012)Jorgensen, Norrman, Gevorgyan, Tromholt,
  Andreasen, and Krebs]{Jorgensen:2011fk}
M.~Jorgensen, K.~Norrman, S.~A. Gevorgyan, T.~Tromholt, B.~Andreasen and F.~C.
  Krebs, \emph{Advanced Materials}, 2012, \textbf{24}, 580--612\relax
\mciteBstWouldAddEndPuncttrue
\mciteSetBstMidEndSepPunct{\mcitedefaultmidpunct}
{\mcitedefaultendpunct}{\mcitedefaultseppunct}\relax
\EndOfBibitem
\bibitem[R{\"o}sch \emph{et~al.}(2012)R{\"o}sch, Tanenbaum, Jorgensen, Seeland,
  B{\"a}renklau, Hermenau, Voroshazi, Lloyd, Galagan, Zimmermann, W{\"u}rfel,
  H{\"o}sel, Dam, Gevorgyan, Kudret, Maes, Lutsen, Vanderzande, Andriessen,
  Teran-Escobar, Lira-Cantu, Rivaton, Uzuno{\u g}lu, Germack, Andreasen,
  Madsen, Norrman, Hoppe, and Krebs]{Rosch:2012fg}
R.~R{\"o}sch, D.~M. Tanenbaum, M.~Jorgensen, M.~Seeland, M.~B{\"a}renklau,
  M.~Hermenau, E.~Voroshazi, M.~T. Lloyd, Y.~Galagan, B.~Zimmermann,
  U.~W{\"u}rfel, M.~H{\"o}sel, H.~F. Dam, S.~A. Gevorgyan, S.~Kudret, W.~Maes,
  L.~Lutsen, D.~Vanderzande, R.~Andriessen, G.~Teran-Escobar, M.~Lira-Cantu,
  A.~Rivaton, G.~Y. Uzuno{\u g}lu, D.~Germack, B.~Andreasen, M.~V. Madsen,
  K.~Norrman, H.~Hoppe and F.~C. Krebs, \emph{Energy {\&} Environmental
  Science}, 2012, \textbf{5}, 6521\relax
\mciteBstWouldAddEndPuncttrue
\mciteSetBstMidEndSepPunct{\mcitedefaultmidpunct}
{\mcitedefaultendpunct}{\mcitedefaultseppunct}\relax
\EndOfBibitem
\bibitem[Andreasen \emph{et~al.}(2012)Andreasen, Tanenbaum, Hermenau,
  Voroshazi, Lloyd, Galagan, Zimmernann, Kudret, Maes, Lutsen, Vanderzande,
  W{\"u}rfel, Andriessen, R{\"o}sch, Hoppe, Teran-Escobar, Lira-Cantu, Rivaton,
  Uzuno{\u g}lu, Germack, H{\"o}sel, Dam, Jorgensen, Gevorgyan, Madsen,
  Bundgaard, Krebs, and Norrman]{Andreasen:2012cy}
B.~Andreasen, D.~M. Tanenbaum, M.~Hermenau, E.~Voroshazi, M.~T. Lloyd,
  Y.~Galagan, B.~Zimmernann, S.~Kudret, W.~Maes, L.~Lutsen, D.~Vanderzande,
  U.~W{\"u}rfel, R.~Andriessen, R.~R{\"o}sch, H.~Hoppe, G.~Teran-Escobar,
  M.~Lira-Cantu, A.~Rivaton, G.~Y. Uzuno{\u g}lu, D.~S. Germack, M.~H{\"o}sel,
  H.~F. Dam, M.~Jorgensen, S.~A. Gevorgyan, M.~V. Madsen, E.~Bundgaard, F.~C.
  Krebs and K.~Norrman, \emph{Physical Chemistry Chemical Physics}, 2012,
  \textbf{14}, 11780\relax
\mciteBstWouldAddEndPuncttrue
\mciteSetBstMidEndSepPunct{\mcitedefaultmidpunct}
{\mcitedefaultendpunct}{\mcitedefaultseppunct}\relax
\EndOfBibitem
\bibitem[Brabec \emph{et~al.}(2010)Brabec, Gowrisanker, Halls, Laird, Jia, and
  Williams]{Brabec:2010hg}
C.~J. Brabec, S.~Gowrisanker, J.~J.~M. Halls, D.~Laird, S.~Jia and S.~P.
  Williams, \emph{Advanced Materials}, 2010, \textbf{22}, 3839--3856\relax
\mciteBstWouldAddEndPuncttrue
\mciteSetBstMidEndSepPunct{\mcitedefaultmidpunct}
{\mcitedefaultendpunct}{\mcitedefaultseppunct}\relax
\EndOfBibitem
\bibitem[Grossiord \emph{et~al.}(2012)Grossiord, Kroon, Andriessen, and
  Blom]{Grossiord:2012jk}
N.~Grossiord, J.~M. Kroon, R.~Andriessen and P.~W. Blom, \emph{Organic
  Electronics}, 2012, \textbf{13}, 432--456\relax
\mciteBstWouldAddEndPuncttrue
\mciteSetBstMidEndSepPunct{\mcitedefaultmidpunct}
{\mcitedefaultendpunct}{\mcitedefaultseppunct}\relax
\EndOfBibitem
\bibitem[Lee \emph{et~al.}(2012)Lee, Jung, Jo, and Jo]{Lee:2012uk}
J.~U. Lee, J.~W. Jung, J.~W. Jo and W.~H. Jo, \emph{Journal of Materials
  Chemistry}, 2012, \textbf{22}, 24265--24283\relax
\mciteBstWouldAddEndPuncttrue
\mciteSetBstMidEndSepPunct{\mcitedefaultmidpunct}
{\mcitedefaultendpunct}{\mcitedefaultseppunct}\relax
\EndOfBibitem
\bibitem[Peters \emph{et~al.}(2011)Peters, Sachs-Quintana, Kastrop,
  Beaupr{\'e}, Leclerc, and Mcgehee]{Peters:2011kd}
C.~H. Peters, I.~T. Sachs-Quintana, J.~P. Kastrop, S.~Beaupr{\'e}, M.~Leclerc
  and M.~D. Mcgehee, \emph{Advanced Energy Materials}, 2011, \textbf{1},
  491--494\relax
\mciteBstWouldAddEndPuncttrue
\mciteSetBstMidEndSepPunct{\mcitedefaultmidpunct}
{\mcitedefaultendpunct}{\mcitedefaultseppunct}\relax
\EndOfBibitem
\bibitem[Yip and Jen(2012)]{Yip:2012it}
H.-L. Yip and A.~K.~Y. Jen, \emph{Energy {\&} Environmental Science}, 2012,
  \textbf{5}, 5994\relax
\mciteBstWouldAddEndPuncttrue
\mciteSetBstMidEndSepPunct{\mcitedefaultmidpunct}
{\mcitedefaultendpunct}{\mcitedefaultseppunct}\relax
\EndOfBibitem
\bibitem[Manceau \emph{et~al.}(2011)Manceau, Bundgaard, Carl{\'e}, Hagemann,
  Helgesen, S{\o}ndergaard, J{\o}rgensen, and Krebs]{Manceau:2011km}
M.~Manceau, E.~Bundgaard, J.~E. Carl{\'e}, O.~Hagemann, M.~Helgesen,
  R.~S{\o}ndergaard, M.~J{\o}rgensen and F.~C. Krebs, \emph{Journal of
  Materials Chemistry}, 2011, \textbf{21}, 4132\relax
\mciteBstWouldAddEndPuncttrue
\mciteSetBstMidEndSepPunct{\mcitedefaultmidpunct}
{\mcitedefaultendpunct}{\mcitedefaultseppunct}\relax
\EndOfBibitem
\bibitem[Tournebize \emph{et~al.}(2012)Tournebize, Bussi{\`e}re,
  Wong-Wah-Chung, Th{\'e}rias, Rivaton, Gardette, Beaupr{\'e}, and
  Leclerc]{Tournebize:2012cl}
A.~Tournebize, P.-O. Bussi{\`e}re, P.~Wong-Wah-Chung, S.~Th{\'e}rias,
  A.~Rivaton, J.-L. Gardette, S.~Beaupr{\'e} and M.~Leclerc, \emph{Advanced
  Energy Materials}, 2012, \textbf{3}, 478--487\relax
\mciteBstWouldAddEndPuncttrue
\mciteSetBstMidEndSepPunct{\mcitedefaultmidpunct}
{\mcitedefaultendpunct}{\mcitedefaultseppunct}\relax
\EndOfBibitem
\bibitem[Manceau \emph{et~al.}(2009)Manceau, Rivaton, Gardette, Guillerez, and
  Lema{\^\i}tre]{Manceau:2009kw}
M.~Manceau, A.~Rivaton, J.-L. Gardette, S.~Guillerez and N.~Lema{\^\i}tre,
  \emph{Polymer Degradation and Stability}, 2009, \textbf{94}, 898--907\relax
\mciteBstWouldAddEndPuncttrue
\mciteSetBstMidEndSepPunct{\mcitedefaultmidpunct}
{\mcitedefaultendpunct}{\mcitedefaultseppunct}\relax
\EndOfBibitem
\bibitem[Hintz \emph{et~al.}(2010)Hintz, Egelhaaf, Peisert, and
  Chass{\'e}]{Hintz:2010eq}
H.~Hintz, H.-J. Egelhaaf, H.~Peisert and T.~Chass{\'e}, \emph{Polymer
  Degradation and Stability}, 2010, \textbf{95}, 818--825\relax
\mciteBstWouldAddEndPuncttrue
\mciteSetBstMidEndSepPunct{\mcitedefaultmidpunct}
{\mcitedefaultendpunct}{\mcitedefaultseppunct}\relax
\EndOfBibitem
\bibitem[Hintz \emph{et~al.}(2011)Hintz, Egelhaaf, L{\"u}er, Hauch, Peisert,
  and Chass{\'e}]{Hintz:2011ds}
H.~Hintz, H.-J. Egelhaaf, L.~L{\"u}er, J.~Hauch, H.~Peisert and T.~Chass{\'e},
  \emph{Chemistry Of Materials}, 2011, \textbf{23}, 145--154\relax
\mciteBstWouldAddEndPuncttrue
\mciteSetBstMidEndSepPunct{\mcitedefaultmidpunct}
{\mcitedefaultendpunct}{\mcitedefaultseppunct}\relax
\EndOfBibitem
\bibitem[Deschler \emph{et~al.}(2012)Deschler, De~Sio, Von~Hauff, Kutka,
  Sauermann, Egelhaaf, Hauch, and Da~Como]{Deschler:2012ci}
F.~Deschler, A.~De~Sio, E.~Von~Hauff, P.~Kutka, T.~Sauermann, H.~J. Egelhaaf,
  J.~Hauch and E.~Da~Como, \emph{Advanced Functional Materials}, 2012,
  \textbf{22}, 1461--1469\relax
\mciteBstWouldAddEndPuncttrue
\mciteSetBstMidEndSepPunct{\mcitedefaultmidpunct}
{\mcitedefaultendpunct}{\mcitedefaultseppunct}\relax
\EndOfBibitem
\bibitem[Abdou and Holdcroft(1993)]{Abdou:1993vr}
M.~S. Abdou and S.~Holdcroft, \emph{Macromolecules}, 1993, \textbf{26},
  2954--2962\relax
\mciteBstWouldAddEndPuncttrue
\mciteSetBstMidEndSepPunct{\mcitedefaultmidpunct}
{\mcitedefaultendpunct}{\mcitedefaultseppunct}\relax
\EndOfBibitem
\bibitem[Abdou and Holdcroft(1995)]{Abdou:1995uu}
M.~Abdou and S.~Holdcroft, \emph{Canadian Journal of Chemistry}, 1995,
  \textbf{73}, 1893--1901\relax
\mciteBstWouldAddEndPuncttrue
\mciteSetBstMidEndSepPunct{\mcitedefaultmidpunct}
{\mcitedefaultendpunct}{\mcitedefaultseppunct}\relax
\EndOfBibitem
\bibitem[Abdou \emph{et~al.}(1997)Abdou, Orfino, Son, and
  Holdcroft]{Abdou:1997ux}
M.~Abdou, F.~Orfino, Y.~Son and S.~Holdcroft, \emph{Journal Of The American
  Chemical Society}, 1997, \textbf{119}, 4518--4524\relax
\mciteBstWouldAddEndPuncttrue
\mciteSetBstMidEndSepPunct{\mcitedefaultmidpunct}
{\mcitedefaultendpunct}{\mcitedefaultseppunct}\relax
\EndOfBibitem
\bibitem[Manceau \emph{et~al.}(2008)Manceau, Rivaton, and
  Gardette]{Manceau:2008kr}
M.~Manceau, A.~Rivaton and J.-L. Gardette, \emph{Macromolecular Rapid
  Communications}, 2008, \textbf{29}, 1823--1827\relax
\mciteBstWouldAddEndPuncttrue
\mciteSetBstMidEndSepPunct{\mcitedefaultmidpunct}
{\mcitedefaultendpunct}{\mcitedefaultseppunct}\relax
\EndOfBibitem
\bibitem[Reese \emph{et~al.}(2008)Reese, Morfa, White, Kopidakis, Shaheen,
  Rumbles, and Ginley]{Reese:2008ey}
M.~O. Reese, A.~J. Morfa, M.~S. White, N.~Kopidakis, S.~E. Shaheen, G.~Rumbles
  and D.~S. Ginley, \emph{Solar Energy Materials and Solar Cells}, 2008,
  \textbf{92}, 746--752\relax
\mciteBstWouldAddEndPuncttrue
\mciteSetBstMidEndSepPunct{\mcitedefaultmidpunct}
{\mcitedefaultendpunct}{\mcitedefaultseppunct}\relax
\EndOfBibitem
\bibitem[Manceau \emph{et~al.}(2010)Manceau, Gaume, Rivaton, Gardette, Monier,
  and Bideux]{Manceau:2010cb}
M.~Manceau, J.~Gaume, A.~Rivaton, J.-L. Gardette, G.~Monier and L.~Bideux,
  \emph{Thin Solid Films}, 2010, \textbf{518}, 7113--7118\relax
\mciteBstWouldAddEndPuncttrue
\mciteSetBstMidEndSepPunct{\mcitedefaultmidpunct}
{\mcitedefaultendpunct}{\mcitedefaultseppunct}\relax
\EndOfBibitem
\bibitem[Norrman \emph{et~al.}(2010)Norrman, Madsen, Gevorgyan, and
  Krebs]{Norrman:2010bq}
K.~Norrman, M.~V. Madsen, S.~A. Gevorgyan and F.~C. Krebs, \emph{Journal Of The
  American Chemical Society}, 2010, \textbf{132}, 16883--16892\relax
\mciteBstWouldAddEndPuncttrue
\mciteSetBstMidEndSepPunct{\mcitedefaultmidpunct}
{\mcitedefaultendpunct}{\mcitedefaultseppunct}\relax
\EndOfBibitem
\bibitem[Reese \emph{et~al.}(2010)Reese, Nardes, Rupert, Larsen, Olson, Lloyd,
  Shaheen, Ginley, Rumbles, and Kopidakis]{Reese:2010ev}
M.~O. Reese, A.~M. Nardes, B.~L. Rupert, R.~E. Larsen, D.~C. Olson, M.~T.
  Lloyd, S.~E. Shaheen, D.~S. Ginley, G.~Rumbles and N.~Kopidakis,
  \emph{Advanced Functional Materials}, 2010, \textbf{20}, 3476--3483\relax
\mciteBstWouldAddEndPuncttrue
\mciteSetBstMidEndSepPunct{\mcitedefaultmidpunct}
{\mcitedefaultendpunct}{\mcitedefaultseppunct}\relax
\EndOfBibitem
\bibitem[Hintz \emph{et~al.}(2011)Hintz, Peisert, Egelhaaf, and
  Chasse]{Hintz:2011kw}
H.~Hintz, H.~Peisert, H.~J. Egelhaaf and T.~Chasse, \emph{Journal Of Physical
  Chemistry C}, 2011, \textbf{115}, 13373--13376\relax
\mciteBstWouldAddEndPuncttrue
\mciteSetBstMidEndSepPunct{\mcitedefaultmidpunct}
{\mcitedefaultendpunct}{\mcitedefaultseppunct}\relax
\EndOfBibitem
\bibitem[Sperlich \emph{et~al.}(2011)Sperlich, Kraus, Deibel, Blok, Schmidt,
  and Dyakonov]{Sperlich:2011bl}
A.~Sperlich, H.~Kraus, C.~Deibel, H.~Blok, J.~Schmidt and V.~Dyakonov,
  \emph{The Journal of Physical Chemistry B}, 2011, \textbf{115},
  13513--13518\relax
\mciteBstWouldAddEndPuncttrue
\mciteSetBstMidEndSepPunct{\mcitedefaultmidpunct}
{\mcitedefaultendpunct}{\mcitedefaultseppunct}\relax
\EndOfBibitem
\bibitem[Seemann \emph{et~al.}(2011)Seemann, Sauermann, Lungenschmied,
  Armbruster, Bauer, Egelhaaf, and Hauch]{Seemann:2011cd}
A.~Seemann, T.~Sauermann, C.~Lungenschmied, O.~Armbruster, S.~Bauer, H.-J.
  Egelhaaf and J.~Hauch, \emph{Solar Energy}, 2011, \textbf{85},
  1238--1249\relax
\mciteBstWouldAddEndPuncttrue
\mciteSetBstMidEndSepPunct{\mcitedefaultmidpunct}
{\mcitedefaultendpunct}{\mcitedefaultseppunct}\relax
\EndOfBibitem
\bibitem[Guerrero \emph{et~al.}(2012)Guerrero, Boix, Marchesi,
  Ripolles-Sanchis, Pereira, and Garcia-Belmonte]{Guerrero:2012dg}
A.~Guerrero, P.~P. Boix, L.~F. Marchesi, T.~Ripolles-Sanchis, E.~C. Pereira and
  G.~Garcia-Belmonte, \emph{Solar Energy Materials and Solar Cells}, 2012,
  \textbf{100}, 185--191\relax
\mciteBstWouldAddEndPuncttrue
\mciteSetBstMidEndSepPunct{\mcitedefaultmidpunct}
{\mcitedefaultendpunct}{\mcitedefaultseppunct}\relax
\EndOfBibitem
\bibitem[Hoke \emph{et~al.}(2012)Hoke, Sachs-Quintana, Lloyd, Kauvar, Mateker,
  Nardes, Peters, Kopidakis, and Mcgehee]{Hoke:2012ec}
E.~T. Hoke, I.~T. Sachs-Quintana, M.~T. Lloyd, I.~Kauvar, W.~R. Mateker, A.~M.
  Nardes, C.~H. Peters, N.~Kopidakis and M.~D. Mcgehee, \emph{Advanced Energy
  Materials}, 2012, \textbf{2}, 1351--1357\relax
\mciteBstWouldAddEndPuncttrue
\mciteSetBstMidEndSepPunct{\mcitedefaultmidpunct}
{\mcitedefaultendpunct}{\mcitedefaultseppunct}\relax
\EndOfBibitem
\bibitem[Hintz \emph{et~al.}(2012)Hintz, Sessler, Peisert, Egelhaaf, and
  Chass{\'e}]{Hintz:2012vk}
H.~Hintz, C.~Sessler, H.~Peisert, H.-J. Egelhaaf and T.~Chass{\'e},
  \emph{Chemistry Of Materials}, 2012, \textbf{24}, 2739--2743\relax
\mciteBstWouldAddEndPuncttrue
\mciteSetBstMidEndSepPunct{\mcitedefaultmidpunct}
{\mcitedefaultendpunct}{\mcitedefaultseppunct}\relax
\EndOfBibitem
\bibitem[Cook \emph{et~al.}(2012)Cook, Furube, and Katoh]{Cook:2012hw}
S.~Cook, A.~Furube and R.~Katoh, \emph{Journal of Materials Chemistry}, 2012,
  \textbf{22}, 4282\relax
\mciteBstWouldAddEndPuncttrue
\mciteSetBstMidEndSepPunct{\mcitedefaultmidpunct}
{\mcitedefaultendpunct}{\mcitedefaultseppunct}\relax
\EndOfBibitem
\bibitem[Distler \emph{et~al.}(2012)Distler, Kutka, Sauermann, Egelhaaf, Guldi,
  Di~Nuzzo, Meskers, and Janssen]{Distler:2012wl}
A.~Distler, P.~Kutka, T.~Sauermann, H.-J. Egelhaaf, D.~M. Guldi, D.~Di~Nuzzo,
  S.~C. Meskers and R.~A. Janssen, \emph{Chemistry Of Materials}, 2012,
  \textbf{24}, 4397--4405\relax
\mciteBstWouldAddEndPuncttrue
\mciteSetBstMidEndSepPunct{\mcitedefaultmidpunct}
{\mcitedefaultendpunct}{\mcitedefaultseppunct}\relax
\EndOfBibitem
\bibitem[Tournebize \emph{et~al.}(2013)Tournebize, Bussi{\`e}re, Rivaton,
  Gardette, Medlej, Hiorns, Dagron-Lartigau, Krebs, and
  Norrman]{Tournebize:2013dy}
A.~Tournebize, P.-O. Bussi{\`e}re, A.~Rivaton, J.-L. Gardette, H.~Medlej, R.~C.
  Hiorns, C.~Dagron-Lartigau, F.~C. Krebs and K.~Norrman, \emph{Chemistry Of
  Materials}, 2013, \textbf{25}, 4522--4528\relax
\mciteBstWouldAddEndPuncttrue
\mciteSetBstMidEndSepPunct{\mcitedefaultmidpunct}
{\mcitedefaultendpunct}{\mcitedefaultseppunct}\relax
\EndOfBibitem
\bibitem[Street \emph{et~al.}(2012)Street, Northrup, and Krusor]{Street:2012kr}
R.~A. Street, J.~E. Northrup and B.~S. Krusor, \emph{Physical Review B}, 2012,
  \textbf{85}, 205211\relax
\mciteBstWouldAddEndPuncttrue
\mciteSetBstMidEndSepPunct{\mcitedefaultmidpunct}
{\mcitedefaultendpunct}{\mcitedefaultseppunct}\relax
\EndOfBibitem
\bibitem[Leung(2013)]{e2}
K.~Leung, \emph{Chemical Physics Letters}, 2013, \textbf{568-569}, 1--8\relax
\mciteBstWouldAddEndPuncttrue
\mciteSetBstMidEndSepPunct{\mcitedefaultmidpunct}
{\mcitedefaultendpunct}{\mcitedefaultseppunct}\relax
\EndOfBibitem
\bibitem[Barnes \emph{et~al.}(2006)Barnes, Hjorth, and
  Mihalopoulos]{Barnes:2006dl}
I.~Barnes, J.~Hjorth and N.~Mihalopoulos, \emph{Chemical Reviews}, 2006,
  \textbf{106}, 940--975\relax
\mciteBstWouldAddEndPuncttrue
\mciteSetBstMidEndSepPunct{\mcitedefaultmidpunct}
{\mcitedefaultendpunct}{\mcitedefaultseppunct}\relax
\EndOfBibitem
\bibitem[Barckholtz \emph{et~al.}(2001)Barckholtz, Barckholtz, and
  Hadad]{Barckholtz:2001wn}
C.~Barckholtz, T.~A. Barckholtz and C.~M. Hadad, \emph{The Journal of Physical
  Chemistry A}, 2001, \textbf{105}, 140--152\relax
\mciteBstWouldAddEndPuncttrue
\mciteSetBstMidEndSepPunct{\mcitedefaultmidpunct}
{\mcitedefaultendpunct}{\mcitedefaultseppunct}\relax
\EndOfBibitem
\bibitem[Song \emph{et~al.}(2012)Song, Fanelli, Cook, Bai, and
  Parish]{Song:2012el}
X.~X. Song, M.~G.~M. Fanelli, J.~M.~J. Cook, F.~F. Bai and C.~A.~C. Parish,
  \emph{The Journal of Physical Chemistry A}, 2012, \textbf{116},
  4934--4946\relax
\mciteBstWouldAddEndPuncttrue
\mciteSetBstMidEndSepPunct{\mcitedefaultmidpunct}
{\mcitedefaultendpunct}{\mcitedefaultseppunct}\relax
\EndOfBibitem
\bibitem[Kresse and Furthmuller(1996)]{Kresse:1996vf}
G.~Kresse and J.~Furthmuller, \emph{Physical Review B}, 1996, \textbf{54},
  11169--11186\relax
\mciteBstWouldAddEndPuncttrue
\mciteSetBstMidEndSepPunct{\mcitedefaultmidpunct}
{\mcitedefaultendpunct}{\mcitedefaultseppunct}\relax
\EndOfBibitem
\bibitem[Perdew \emph{et~al.}(1996)Perdew, Burke, and Ernzerhof]{Perdew:1996ug}
J.~Perdew, K.~Burke and M.~Ernzerhof, \emph{Physical Review Letters}, 1996,
  \textbf{77}, 3865--3868\relax
\mciteBstWouldAddEndPuncttrue
\mciteSetBstMidEndSepPunct{\mcitedefaultmidpunct}
{\mcitedefaultendpunct}{\mcitedefaultseppunct}\relax
\EndOfBibitem
\bibitem[Perdew \emph{et~al.}(1996)Perdew, Emzerhof, and Burke]{Perdew:1996tq}
J.~Perdew, M.~Emzerhof and K.~Burke, \emph{Journal Of Chemical Physics}, 1996,
  \textbf{105}, 9982--9985\relax
\mciteBstWouldAddEndPuncttrue
\mciteSetBstMidEndSepPunct{\mcitedefaultmidpunct}
{\mcitedefaultendpunct}{\mcitedefaultseppunct}\relax
\EndOfBibitem
\bibitem[Adamo and Barone(1999)]{Adamo:1999hv}
C.~Adamo and V.~Barone, \emph{Journal Of Chemical Physics}, 1999, \textbf{110},
  6158\relax
\mciteBstWouldAddEndPuncttrue
\mciteSetBstMidEndSepPunct{\mcitedefaultmidpunct}
{\mcitedefaultendpunct}{\mcitedefaultseppunct}\relax
\EndOfBibitem
\bibitem[Henkelman \emph{et~al.}(2000)Henkelman, Uberuaga, and
  Jonsson]{Henkelman:2000ez}
G.~Henkelman, B.~P. Uberuaga and H.~Jonsson, \emph{The Journal of Chemical
  Physics}, 2000, \textbf{113}, 9901\relax
\mciteBstWouldAddEndPuncttrue
\mciteSetBstMidEndSepPunct{\mcitedefaultmidpunct}
{\mcitedefaultendpunct}{\mcitedefaultseppunct}\relax
\EndOfBibitem
\bibitem[Henkelman and Jonsson(1999)]{Henkelman:1999vr}
G.~Henkelman and H.~Jonsson, \emph{Journal Of Chemical Physics}, 1999,
  \textbf{111}, 7010--7022\relax
\mciteBstWouldAddEndPuncttrue
\mciteSetBstMidEndSepPunct{\mcitedefaultmidpunct}
{\mcitedefaultendpunct}{\mcitedefaultseppunct}\relax
\EndOfBibitem
\bibitem[Arosio \emph{et~al.}(2008)Arosio, Moreno, Famulari, Raos, Catellani,
  and Meille]{Arosio:2008tw}
P.~Arosio, M.~Moreno, A.~Famulari, G.~Raos, M.~Catellani and S.~Meille,
  \emph{Chemistry Of Materials}, 2008, \textbf{21}, 78--87\relax
\mciteBstWouldAddEndPuncttrue
\mciteSetBstMidEndSepPunct{\mcitedefaultmidpunct}
{\mcitedefaultendpunct}{\mcitedefaultseppunct}\relax
\EndOfBibitem
\bibitem[Frisch \emph{et~al.}()Frisch, Trucks, Schlegel, Scuseria, Robb,
  Cheeseman, Scalmani, Barone, Mennucci, Petersson, Nakatsuji, Caricato, Li,
  Hratchian, Izmaylov, Bloino, Zheng, Sonnenberg, Hada, Ehara, Toyota, Fukuda,
  Hasegawa, Ishida, Nakajima, Honda, Kitao, Nakai, Vreven, Montgomery, Peralta,
  Ogliaro, Bearpark, Heyd, Brothers, Kudin, Staroverov, Kobayashi, Normand,
  Raghavachari, Rendell, Burant, Iyengar, Tomasi, Cossi, Rega, Millam, Klene,
  Knox, Cross, Bakken, Adamo, Jaramillo, Gomperts, Stratmann, Yazyev, Austin,
  Cammi, Pomelli, Ochterski, Martin, Morokuma, Zakrzewski, Voth, Salvador,
  Dannenberg, Dapprich, Daniels, Farkas, Foresman, Ortiz, Cioslowski, and
  Fox]{g09}
M.~J. Frisch, G.~W. Trucks, H.~B. Schlegel, G.~E. Scuseria, M.~A. Robb, J.~R.
  Cheeseman, G.~Scalmani, V.~Barone, B.~Mennucci, G.~A. Petersson,
  H.~Nakatsuji, M.~Caricato, X.~Li, H.~P. Hratchian, A.~F. Izmaylov, J.~Bloino,
  G.~Zheng, J.~L. Sonnenberg, M.~Hada, M.~Ehara, K.~Toyota, R.~Fukuda,
  J.~Hasegawa, M.~Ishida, T.~Nakajima, Y.~Honda, O.~Kitao, H.~Nakai, T.~Vreven,
  J.~A. Montgomery, {Jr.}, J.~E. Peralta, F.~Ogliaro, M.~Bearpark, J.~J. Heyd,
  E.~Brothers, K.~N. Kudin, V.~N. Staroverov, R.~Kobayashi, J.~Normand,
  K.~Raghavachari, A.~Rendell, J.~C. Burant, S.~S. Iyengar, J.~Tomasi,
  M.~Cossi, N.~Rega, J.~M. Millam, M.~Klene, J.~E. Knox, J.~B. Cross,
  V.~Bakken, C.~Adamo, J.~Jaramillo, R.~Gomperts, R.~E. Stratmann, O.~Yazyev,
  A.~J. Austin, R.~Cammi, C.~Pomelli, J.~W. Ochterski, R.~L. Martin,
  K.~Morokuma, V.~G. Zakrzewski, G.~A. Voth, P.~Salvador, J.~J. Dannenberg,
  S.~Dapprich, A.~D. Daniels, ….~Farkas, J.~B. Foresman, J.~V. Ortiz,
  J.~Cioslowski and D.~J. Fox, \emph{Gaussian~09 {R}evision {C}.01}, Gaussian
  Inc. Wallingford CT 2009\relax
\mciteBstWouldAddEndPuncttrue
\mciteSetBstMidEndSepPunct{\mcitedefaultmidpunct}
{\mcitedefaultendpunct}{\mcitedefaultseppunct}\relax
\EndOfBibitem
\bibitem[Grimme \emph{et~al.}(2011)Grimme, Ehrlich, and Goerigk]{Grimme:2011cl}
S.~Grimme, S.~Ehrlich and L.~Goerigk, \emph{Journal of computational
  chemistry}, 2011, \textbf{32}, 1456--1465\relax
\mciteBstWouldAddEndPuncttrue
\mciteSetBstMidEndSepPunct{\mcitedefaultmidpunct}
{\mcitedefaultendpunct}{\mcitedefaultseppunct}\relax
\EndOfBibitem
\bibitem[Scalmani and Frisch(2010)]{Scalmani:2010el}
G.~Scalmani and M.~J. Frisch, \emph{The Journal of Chemical Physics}, 2010,
  \textbf{132}, 114110\relax
\mciteBstWouldAddEndPuncttrue
\mciteSetBstMidEndSepPunct{\mcitedefaultmidpunct}
{\mcitedefaultendpunct}{\mcitedefaultseppunct}\relax
\EndOfBibitem
\bibitem[Gross \emph{et~al.}(2004)Gross, Barnes, S{\o}rensen, Kongsted, and
  Mikkelsen]{Gross:2004wx}
A.~Gross, I.~Barnes, R.~M. S{\o}rensen, J.~Kongsted and K.~V. Mikkelsen,
  \emph{The Journal of Physical Chemistry A}, 2004, \textbf{108},
  8659--8671\relax
\mciteBstWouldAddEndPuncttrue
\mciteSetBstMidEndSepPunct{\mcitedefaultmidpunct}
{\mcitedefaultendpunct}{\mcitedefaultseppunct}\relax
\EndOfBibitem
\bibitem[Rabek(1996)]{Rabek:1996jx}
J.~F. Rabek, \emph{{Photodegradation of Polymers}}, Springer Berlin Heidelberg,
  Berlin, Heidelberg, 1996\relax
\mciteBstWouldAddEndPuncttrue
\mciteSetBstMidEndSepPunct{\mcitedefaultmidpunct}
{\mcitedefaultendpunct}{\mcitedefaultseppunct}\relax
\EndOfBibitem
\bibitem[Z{\'a}dor \emph{et~al.}(2011)Z{\'a}dor, Taatjes, and
  Fernandes]{Zador:2011kz}
J.~Z{\'a}dor, C.~A. Taatjes and R.~X. Fernandes, \emph{Progress in Energy and
  Combustion Science}, 2011, \textbf{37}, 371--421\relax
\mciteBstWouldAddEndPuncttrue
\mciteSetBstMidEndSepPunct{\mcitedefaultmidpunct}
{\mcitedefaultendpunct}{\mcitedefaultseppunct}\relax
\EndOfBibitem
\bibitem[Audouin \emph{et~al.}(1994)Audouin, Langlois, Verdu, and
  Debruijn]{Audouin:1994tu}
L.~Audouin, V.~Langlois, J.~Verdu and J.~Debruijn, \emph{Journal Of Materials
  Science}, 1994, \textbf{29}, 569--583\relax
\mciteBstWouldAddEndPuncttrue
\mciteSetBstMidEndSepPunct{\mcitedefaultmidpunct}
{\mcitedefaultendpunct}{\mcitedefaultseppunct}\relax
\EndOfBibitem
\bibitem[Cohen \emph{et~al.}(2012)Cohen, Mori-S{\'a}nchez, and
  Yang]{Cohen:2012fm}
A.~J. Cohen, P.~Mori-S{\'a}nchez and W.~Yang, \emph{Chem Rev}, 2012,
  \textbf{112}, 289--320\relax
\mciteBstWouldAddEndPuncttrue
\mciteSetBstMidEndSepPunct{\mcitedefaultmidpunct}
{\mcitedefaultendpunct}{\mcitedefaultseppunct}\relax
\EndOfBibitem
\bibitem[Cohen \emph{et~al.}(2008)Cohen, Mori-Sanchez, and Yang]{Cohen:2008wh}
A.~Cohen, P.~Mori-Sanchez and W.~Yang, \emph{Science}, 2008, \textbf{321},
  792\relax
\mciteBstWouldAddEndPuncttrue
\mciteSetBstMidEndSepPunct{\mcitedefaultmidpunct}
{\mcitedefaultendpunct}{\mcitedefaultseppunct}\relax
\EndOfBibitem
\bibitem[not()]{note}
We further explored the influence of DFT methods on the reaction barriers by
  using hybrid PBEh functional with 50\% of exact exchange and found a reaction
  barrier of 1.43 eV. The 50\% hybrid functional has been reported to have a
  mean error of $-$0.017 eV for the BH42 set of gas phase reaction
  barriers.\cite{Vydrov:2006gq} Systematic comparison of barrier heights with
  different fractions of HF has not been reported for solid state materials in
  the literature. We consider this barrier height of 1.43 eV to be an upper
  bound for the reaction barrier in Eq.~(\ref{eq:ROO}). Once we take into
  account of the 30\% drop of reaction barrier in using a doubled unit cell and
  the entropy correction (which can further reduce the barrier height by up to
  0.1 eV), the ROO reation barrier remains below 1.0 eV.\relax
\mciteBstWouldAddEndPunctfalse
\mciteSetBstMidEndSepPunct{\mcitedefaultmidpunct}
{}{\mcitedefaultseppunct}\relax
\EndOfBibitem
\bibitem[Vydrov \emph{et~al.}(2006)Vydrov, Heyd, Krukau, and
  Scuseria]{Vydrov:2006gq}
O.~A. Vydrov, J.~Heyd, A.~V. Krukau and G.~E. Scuseria, \emph{The Journal of
  Chemical Physics}, 2006, \textbf{125}, 074106\relax
\mciteBstWouldAddEndPuncttrue
\mciteSetBstMidEndSepPunct{\mcitedefaultmidpunct}
{\mcitedefaultendpunct}{\mcitedefaultseppunct}\relax
\EndOfBibitem
\end{mcitethebibliography}
\bibliographystyle{rsc} 
}

\end{document}